\documentclass{iau}
\usepackage{graphicx,natbib}
 
\newcommand{\apj}{ApJ}           
\newcommand{\mnras}{MNRAS}       
\newcommand{\nat}{Nature}
\newcommand{\aap}{A\&A}

\newcommand{\aj}{AJ}
\newcommand{\pasa}{PASA}

\newcommand{\arcsec}{\mbox{\ensuremath{^{\prime\prime}}}}

\title{The Hector Survey: integral field spectroscopy of 100,000 galaxies}

\author[J.\ Bland-Hawthorn]{J.\ Bland-Hawthorn$^1$}

\affiliation{$^1$Sydney Institute for Astronomy (SIfA), School of Physics, The University of Sydney, NSW 2006, Australia
\\email: {\tt jbh@physics.usyd.edu.au}}
\pubyear{2014}
\volume{309}
\jname{Galaxies in 3D across the Universe}
\editors{B. L. Ziegler, F. Combes, H. Dannerbauer, M. Verdugo, eds.}

\begin{document}

\maketitle

\begin{abstract}
In March 2013, the Sydney--AAO Multi-object Integral field spectrograph (SAMI) began a major survey of 3400 galaxies at the AAT, the 
largest of its kind to date. At the time of writing, over a third of the targets have been observed and the scientific impact has
been immediate. The Manga galaxy survey has now started at the SDSS telescope and will target an even 
larger sample of nearby galaxies. In Australia, the community is now gearing up to deliver a major new facility called Hector
that will allow integral field spectroscopy of 100 galaxies observed simultaneously. By the close of the decade, it will
be possible to obtain integral field spectroscopy of 100,000 galaxies over 3000 square degrees of sky
down to $r=17$ (median).
Many of these objects will have HI imaging from the new ASKAP radio surveys.
We discuss the motivation for such a survey and the use of new cosmological simulations that 
are properly matched to the integral field observations. The Hector survey will open up a new and 
unique parameter space for galaxy evolution studies.

\keywords{galaxies: evolution -- galaxies: kinematics and dynamics -- galaxies: structure -- techniques: imaging spectroscopy}
\end{abstract}

\firstsection
\section{Introduction}
A chain of argument is only as strong as its weakest link. In this
respect, our understanding of how galaxies evolve from the first seeds 
to the present day is seriously incomplete. We have a reasonable
picture of how the first dark matter structures came together out of
the initial matter perturbations. But just how gas settled into these
structures to form the first stars and subsequent generations remains
an extremely difficult problem.  From an observational perspective, the 
main advances in evolutionary studies have come from spatially
resolved imaging and spectroscopy across many wavebands, from the development of 
optical and imaging spectroscopy since the 1960s,
the HI and radio continuum surveys of the 1970s, improvements in
infrared imaging and spectroscopy in the 1980s, and the emergence of
mid-infrared, UV, x-ray and gamma-ray satellites in the 1990s to the present 
day. These advances have come from a combination of detailed studies
of individual objects and large surveys of global parameters.
 
Over the past twenty years, imaging surveys from the Hubble Space Telescope
(far field) and the Sloan Digital Sky Survey (near field) have been particularly 
effective in identifying evolution of galaxy parameters with cosmic time and
with environment across large-scale structure. This has been matched by 
large galaxy surveys using multi-object spectroscopy (MOS), most notably at the AAT,
SDSS, VLT, Magellan and Keck telescopes (e.g. \citealt{York00}; \citealt{Colless01};
\citealt{Driver11}). MOS instruments provide a single spectrum
within a fixed fibre aperture at the centre of each galaxy; spatial information must be drawn 
from multi-wavelength broadband images. While this approach has its place, the question
always remained as to whether imaging spectroscopy was feasible for a 
cosmologically significant sample.

In the 1980s, the scientific potential for imaging spectroscopy was 
clear from early Fabry-Perot (\citealt{Bland87,Cecil88,Amram92}) and 
tuneable filter studies (\citealt{Jones01}). After the pioneering work of 
Court\`{e}s et al (1988), integral field spectroscopy (IFS) soon exploited the plunging costs of 
large-area detectors to dominate extragalactic studies today (\citealt{Hill14}).
Recent integral field surveys involving hundreds of nearby galaxies include ATLAS-3D 
\citep{Cappellari11} and CALIFA \citep{Sanchez12}. A useful summary of recent scientific 
results is given by \citet{Glazebrook13}.

So how do we combine the power of the MOS and IFS techniques, the `missing link' in astronomical 
instrumentation? VLT KMOS provides an elegant solution involving small image slicers: this instrument employs 24 
configurable arms that position pickoff mirrors at specified locations in the Nasmyth focal plane. But this expensive technology would be prohibitively costly to adapt for the degree-scale fields of 
cosmologically motivated MOS instruments.
Seven years ago, we began to look at compact fibre bundles (hexabundles) that would work
with existing robotic positioners (\citealt{BlandHawthorn11,Bryant11}). This led to the Sydney--AAO Multi-object Integral field spectrograph (SAMI), the first instrument of its kind. SAMI deploys 13 IFUs, each with a field of view of 15\arcsec\, across a 1-degree patrol field \citep{Croom12}. Each IFU consists of a bundle of 61 optical fibres lightly fused to have a high ($\sim$75\%) filling factor \citep{Bryant14a}. SAMI is installed on the 3.9-m Anglo-Australian Telescope (AAT), feeding the existing AAOmega spectrograph. For the first time, SAMI allows very large samples of IFS observations to be obtained in a short period of time.

\begin{figure}
\centerline{
\includegraphics[scale=0.4]{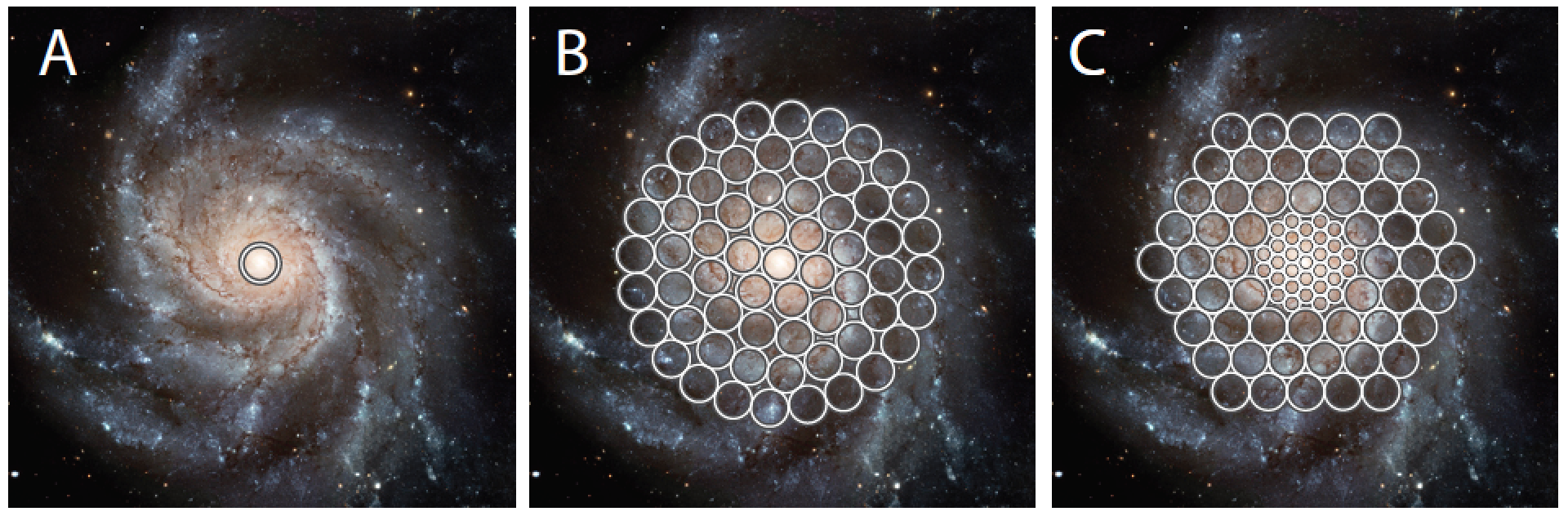}
}
\caption{(a) Single fibre apertures are the basis for galaxy surveys to date; (b) the
61-fibre configuration of the SAMI hexabundle; (c) the 85-fibre (54 large, 31 small)
configuration of the
Hector hexabundle under development at the University of Sydney.
\label{f:bundles}
} 
\end{figure}

\begin{figure}
\centerline{
\includegraphics[scale=0.5]{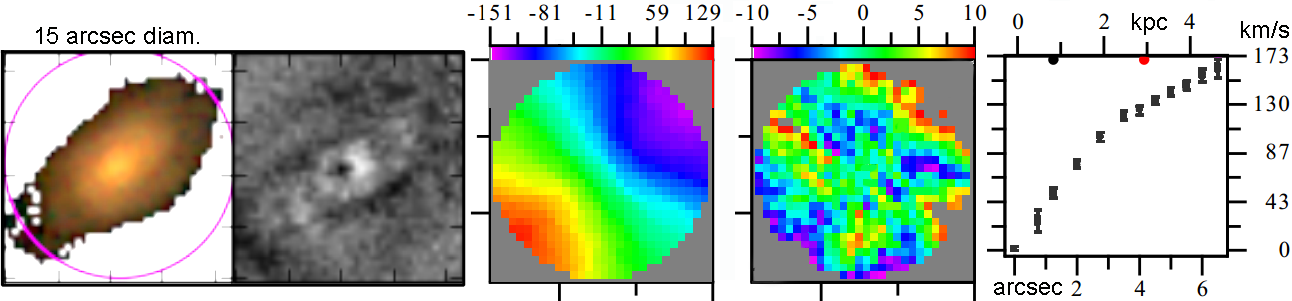}
}
\caption{SAMI results on a nearby galaxy: (a) red stellar continuum; (b) residual after disk model subtracted
to reveal spiral arms; (c) gas velocity field in km s$^{-1}$; (d) $<$7\% residuals after rotation model subtracted; (e)
gas rotation as a function of radius (with errors shown). SAMI also provides stellar velocities and emission-line
ratios giving information on gas chemistry and ionization conditions. The circular bundle in (a) is
dithered using small shifts on sky to remove the footprint (Fig.~\ref{f:bundles}) and fill out the field with complete data.
\label{f:sami}
}
\end{figure}

\begin{figure}
\centerline{
\includegraphics[scale=0.27]{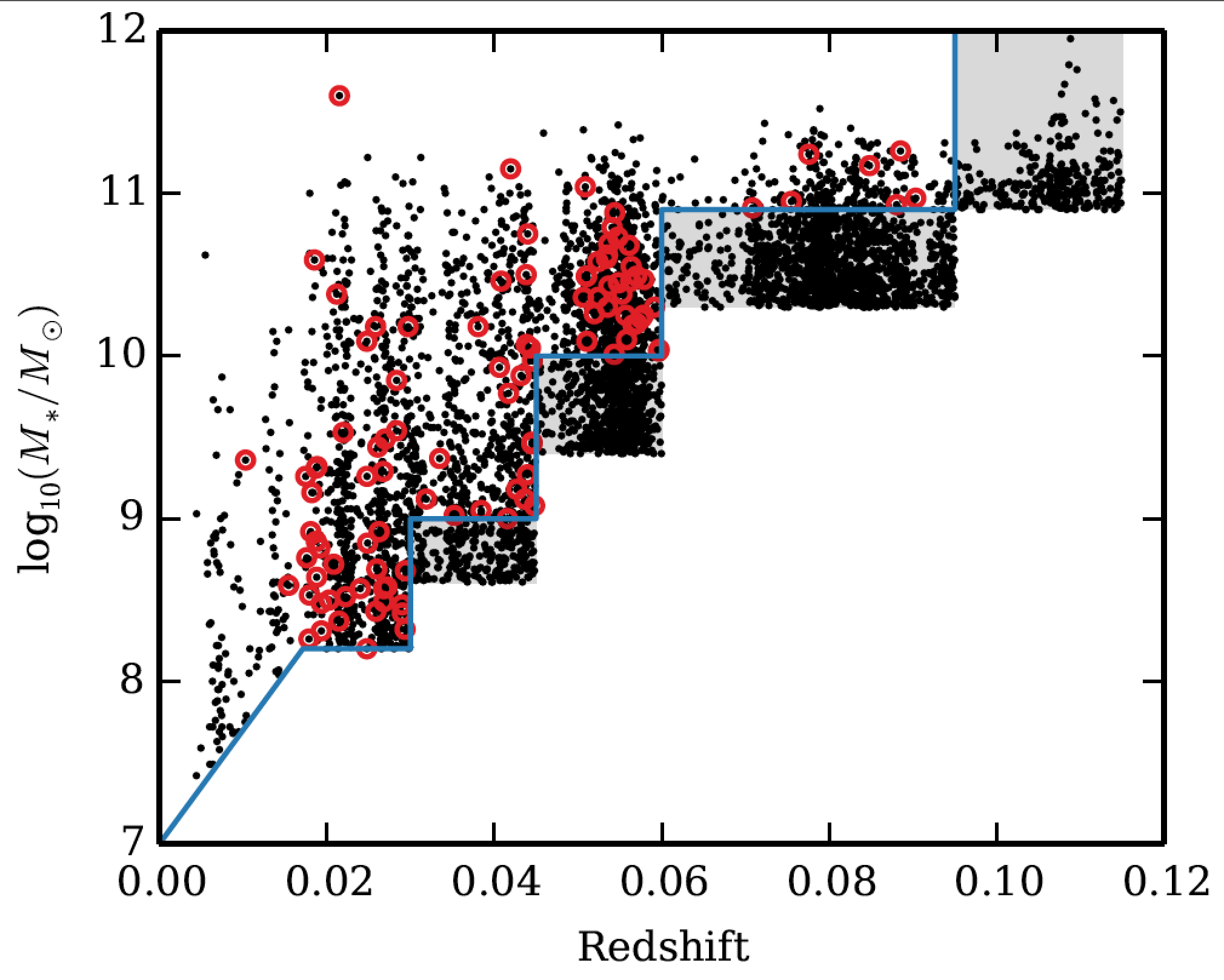}
}
\caption{Stellar mass and redshift for all galaxies in the GAMA field
regions of the SAMI Galaxy Survey (black points); the early data release 
sample are also shown (red circles). The blue boundaries indicate the primary selection criteria,
while the shaded regions indicate lower-priority targets. Large-scale
structure within the GAMA regions is seen in the overdensities of galaxies
at particular redshifts (\citealt{Allen14}).
\label{f:survey}
}
\end{figure}

\section{The SAMI Galaxy Survey: early results}

The SAMI Galaxy Survey is an ongoing project to obtain integral field spectroscopic data for $\sim$3400 galaxies, with an expected completion date of mid 2016. The bundles are dithered with respect to sky
to wash out the footprint (Fig.~\ref{f:bundles}(b)). An example of the data quality is shown in 
Fig.~\ref{f:sami}. At the time of writing, roughly 1000 galaxies have been observed, including the pilot 
survey of $\sim$100 galaxies. The updated instrument performance, target selection, survey parameters and reduction procedures are discussed elsewhere \citep{Bryant14b,Sharp14}. The Galaxy Survey, which includes both cluster and field objects selected from the GAMA survey, had an early data release in 2014 \citep{Allen14}. The use of GAMA fields means that the SAMI data are 
complemented by many surveys, including GALEX MIS, VST KiDS, VISTA VIKING, WISE, 
Herschel-ATLAS, GMRT and ASKAP \citep{Driver11}. 
Early science results include studies of galactic winds \citep{Fogarty12,Ho14}, 
a universal dynamical relation for all galaxy types \citep{Cortese14},
the kinematic morphology--density relation for early-type galaxies \citep{Fogarty14}, 
bar streaming in spiral galaxies \citep{Cecil14}, and
star formation in dwarf galaxies \citep{Richards14}; see the related contributions in this volume.

\section{Hector instrument}

The main limitation of SAMI is the need to plug each bundle into pre-cut metal 
plates. A. Bauer has made a short movie $-$ www.youtube.com/watch?v=MB5H-5XZ9ZE $-$ of 
this labour-intensive process, something we would rather avoid with 100,000 targets in mind. 
The Hector concept (\citealt{Lawrence12}) will exploit the full two-degree field of the AAT prime focus 
using robotically positioned hexabundles.  The AAO has invested several decades in robotic 
technology for positioning fibres over large focal planes. Their most recent
development is the semi-autonomous ``starbug'' that positions fibres (or bundles) with a 
piezo-electric tube that walks over a glass plate. For a movie of how this works in
practice, see here: www.youtube.com/watch?v=YvZDF54Si5w. Recent tests by S. Richards
and C. Betters show that this technology is ideal for moving the bundles around \citep{Goodwin10}.

Much like HETDEX and VLT MUSE before us, we will need to replicate cheap spectrographs
to handle 8500 fibres from the 100 hexabundles. Replicating spectrographs at $R=2000$ is 
relatively inexpensive using 4000-pixel Andor detectors, for example. We are exploring simple 
designs involving curved VPH gratings but it is unclear if this technology
can reach our goal of $R = \lambda/\delta\lambda \approx 4000$ across the optical band
(370-900 nm). A resolution and rough costings should be ready by early 2015. The modular
design of Hector means that the first devices will be ready for use by the end of the SAMI
Galaxy Survey in mid 2016.

The need for a new hexabundle with a supersampled inner bundle (Fig.~\ref{f:bundles}(c)) is to 
minimise loss of spatial resolution over the core regions of galaxies. Essentially all SAMI galaxies are 
brightest over the inner scale length where the gravitational potential is changing most rapidly.
The benefits of proper sampling includes more accurate modelling of the inner potential (e.g.
bars), sensitivity to kinematic substructure, higher order moments, and so forth (\citealt{Naab13}). 
The new design allows for excellent sampling within an effective radius ($R_e$) for most galaxies 
while allowing us to reach $2R_e$ with the larger outer fibre apertures.
This design also has the added benefit of providing better intrinsic sampling of 
more distant and compact targets.

\section{Hector science}

\subsection{Big questions}

\noindent{\it Fossil signatures.}
The formation history of galaxies leaves its imprint on the gas and
stellar kinematic properties of present day galaxies. Epochs dominated
by gas dissipation will result in the formation of flattened stellar
distributions (disks), supported by rotation. During merger dominated
phases the stellar systems experience stripping and violent
relaxation, existing cold gas might be driven to the central regions
causing starbursts, trigger the formation and growth of supermassive
black holes or be expelled form the systems in a galactic wind. This,
in turn, will impact the distribution of cold gas and the kinematics
of forming stars.

Using cosmological `zoom' simulations of representative synthetic
galaxies, \citet{Naab13} demonstrate that 
gas dissipation and merging result in observable features (at the present day) in 
the 2D kinematic properties of galaxies, which are clear
signatures of distinct formation processes. Dissipation favours the
formation of fast rotating systems and line-of-sight velocity
distributions with steep leading wings, a property that can be
directly traced back to the orbital composition of the systems
(\citealt{Rottgers14,Sharma12}). Merging and accretion
can result in fast or slowly rotating systems with counter-rotating
cores, cold nuclear or extended (sometimes counter-rotating) disks
showing dumpbell-like features - all observed in real galaxies. The
strength of these kinematic signatures will be influenced by feedback
from supernovae and AGN, but also by the mass of the galaxies and their
environment.

Low mass star forming disk galaxies favor low-density environments,
predominantly grown by accretion of gas and subsequent in-situ star
formation, and are affected by stellar feedback. Higher mass
early-type galaxies form in high-density environments - possibly
affected by feedback from accreting super-massive black holes - and
their late assembly involves merging with other galaxies, which might
also be of an early type. Before now, it was not possible to perform a
statistically meaningful comparison of kinematic properties of galaxy
populations to observed population properties, like the observed
increasing fraction of slow versus fast rotators for early-type
galaxies as a function of environmental density (\citealt{Cappellari12}).
With the new simulations (e.g. \citealt{Schaye14}), we can tackle these 
questions for the first time using the
large simulated volume and the higher spatial resolution.

With the Hector instrument, the 2D kinematic (gas, stars) and
abundance patterns will be compared over different environs. Preliminary
work will be possible with SAMI and Manga, but the higher target density
across large-scale structure will allow a more detailed statistical 
connection to investigate trends with environment. We will attempt to
(i) identify characteristic formation histories,
(ii) identify properties of progenitor galaxy populations,
(iii) assess the impact of the major feedback mechanisms (from
massive stars and AGN) on the kinematic properties of high and low
mass galaxies. This study that will be supported by
higher resolution cosmological zoom simulations for characteristic
cases.

\begin{figure}
\centerline{
\includegraphics[scale=0.27]{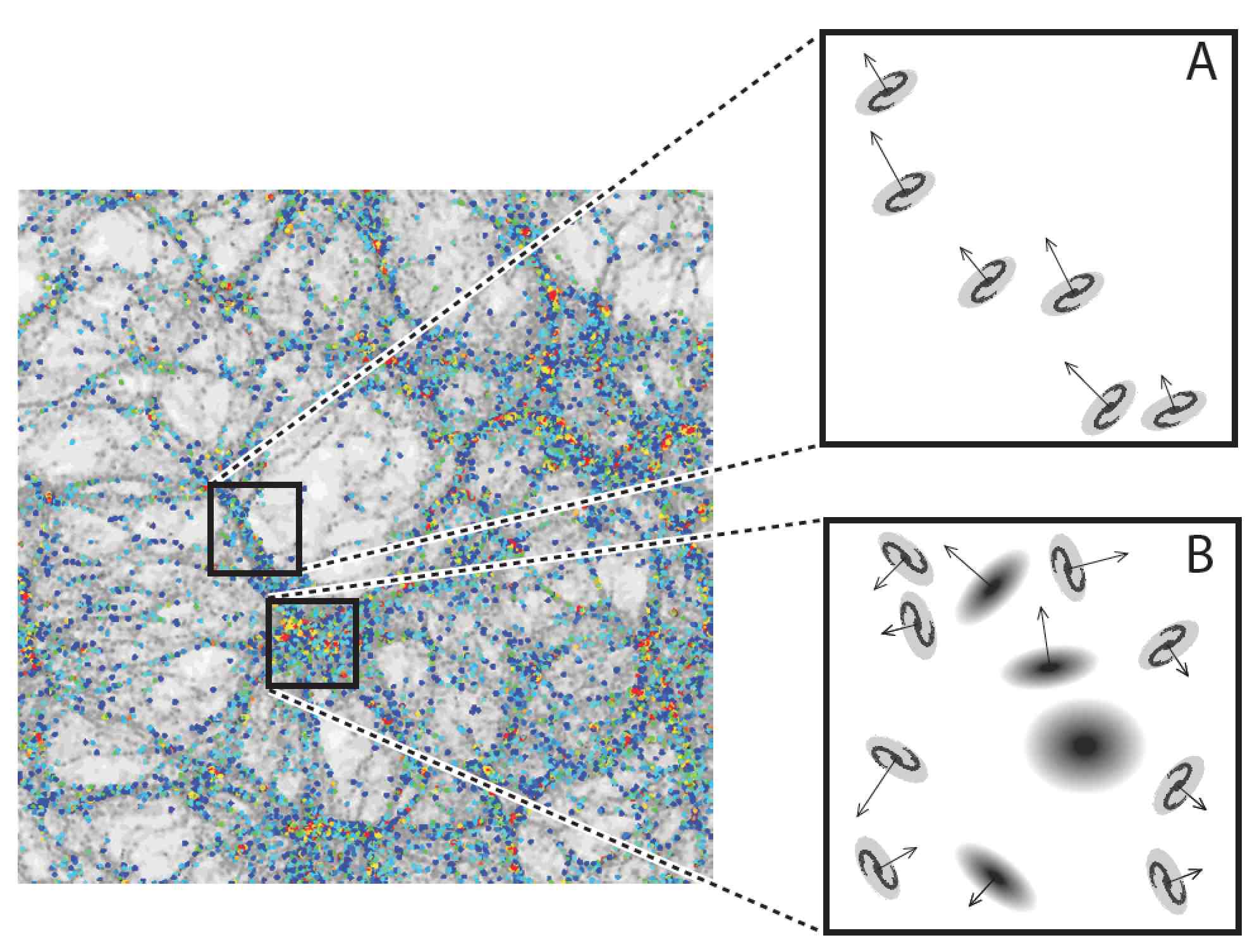}
}
\caption{
Cosmological simulation (GIF consortium) of the universe today within a 30 Mpc volume.
Blue, yellow and red dots are young, middle-aged and old galaxies; the grey is unprocessed
gas. The most detailed simulations predict that (A) low-mass galaxy spins align along 
filaments, and (B) spins are more randomised in dense regions (e.g Dubois et al 2014).
\label{f:GIF}
}
\bigskip
\end{figure}

\begin{figure}
\centerline{
\includegraphics[scale=0.21]{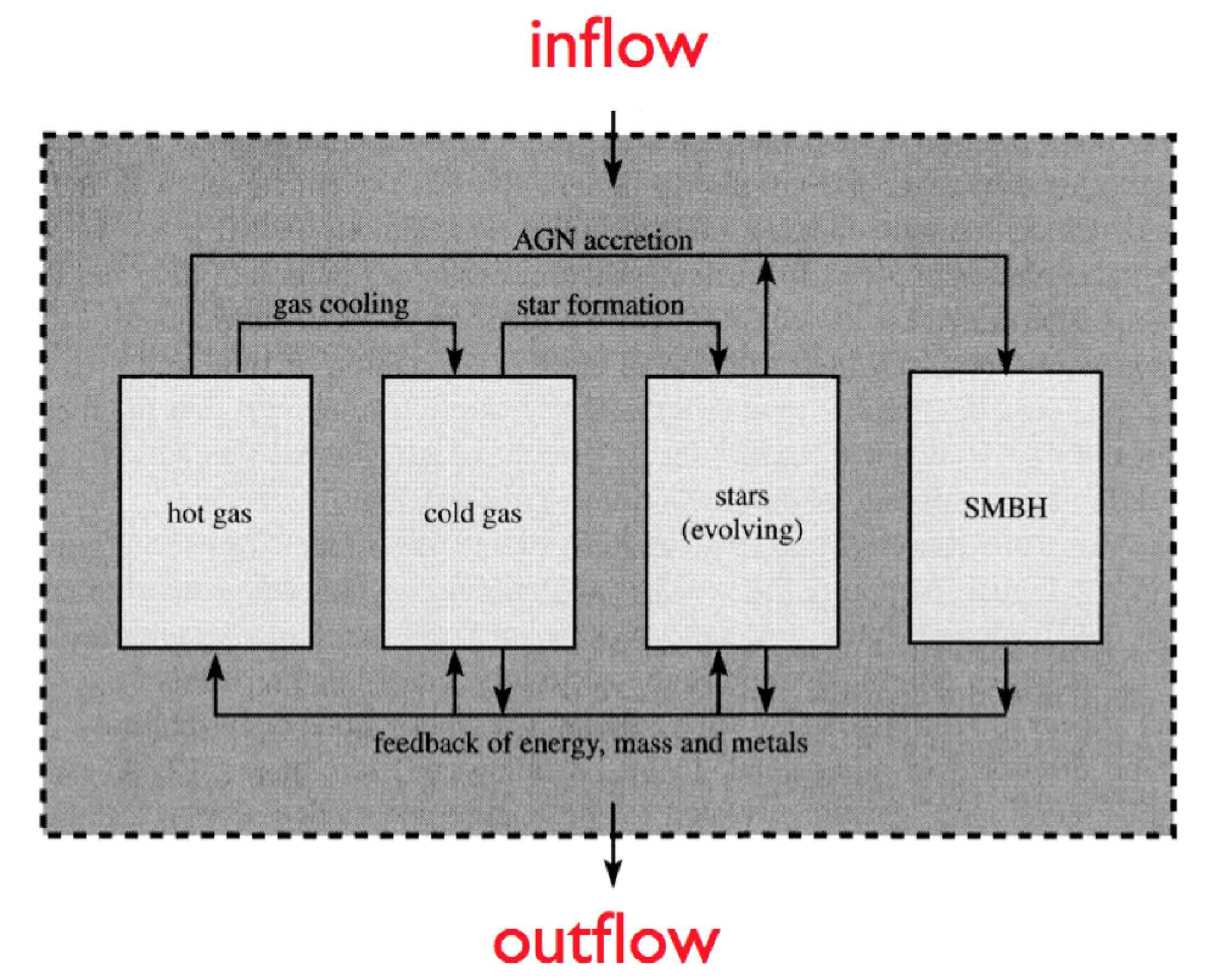}
}
\caption{
A schematic of the inner workings of a galaxy (adapted from \citealt{Mo10}),
akin to a 4-stroke engine fuelled by gas with a complex exhaust system. 
Galaxy evolution is an extremely complex process that {\it demands} much 
larger 3D spectroscopic surveys and the measurement of new parameters.
\label{f:mo1}
}
\bigskip
\end{figure}

\smallskip
\noindent{\it The origin of angular momentum.}
An interesting future area of study is the build-up of angular momentum in galaxies
and across different environs (Fig.~\ref{f:GIF}).
\citet{Codis12} exploit the {\it Horizon} simulations ($z>1.3$) and find an interesting
interplay between the gas and dark matter during accretion onto galaxies. 
These phases separate when sheets 
and filaments form between expanding voids of different sizes. Material
flows out of voids at a rate that depends on the size of the void. The dark matter collapses to form asymmetric sheet or filament but the gas does not. The gas undergoes shocks and compresses to 
form a medium whose
main concentration is offset from the dark matter Ð the protocloud. 
The protocloud oscillates slowly in the gravity
field of the collapsed dark matter sheet and rains onto galaxies as a ÒcoldÓ or a ÒwarmÓ flow. 
This idea has far-reaching
consequences. The wider the separation of gas and dark matter, 
the longer the delay in accreting to the outer
disk leading to galaxies with different spin properties and star formation histories.
The {\it Horizon} team find that low mass
gas-accreting galaxies develop spins aligned
with filaments (Fig.~\ref{f:GIF}) whereas the high mass galaxies are anti-aligned due to major collisions 
\citet{Dubois14}. The signal is expected in the angular momentum vector field (spin direction and magnitude)  when averaged over many thousands of galaxies.

To see this weak signature at $z\sim$ 0.1, we
need to observe of order 60,000 galaxies in a contiguous volume (\citealt{Trowland13}) although
a larger sample may be required (\citealt{Dubois14}).
This requires mass and rotation estimates for stars and/or gas in each galaxy, which Hector is uniquely situated to do.
In this paradigm, the collective properties of galaxies in sheets and filaments may be different in character across the
hierarchy. Once again, 
galaxies within a single filament, for example, are expected to have correlated properties
but the properties can vary between different filaments. If this is correct, it will radically overhaul our ideas on how
galaxies form and evolve because now the disk properties of galaxies are linked to the large-scale environment. In the
earlier ÒclassicalÓ paradigm, shock heating removes any vestige of where the gas came from, thereby removing any
large-scale correlations; the hot gas simply cools and collapses to a disk within the halo (\citealt{Fall80}).

\subsection{Matched data from GAMA and HI surveys}

The GAMA survey is the ideal input catalogue for the SAMI survey because of its supporting
data. As we have shown, galaxies are observed in volume-limited regions at several different 
redshifts (Fig.~\ref{f:survey}) with a Petrosian limit of about $r\approx 17-17.5$ for the faintest sources. 
The GAMA field selection means that our survey data are
complemented by many surveys from UV to radio wavelengths \citep{Driver11}.
Ideally, for the Hector 
survey, we would require the GAMA survey to be extended about a magnitude deeper 
and over a larger region of sky.
By the end of the decade, the Large Synoptic Survey Telescope (LSST) will provide 
very deep multiband photometry across the Southern Sky. 

The GAMA depth and selection complements the all-sky HI surveys soon to be carried out by 
ASKAP in Australia
(\citealt{Duffy12}). Some fraction of the envisaged Hector sources will have spatially resolved (10$^{\prime\prime}$ FWHM) HI imaging and spectroscopy over roughly $\sim 1^{\prime}$ diameter. There are to be 
two major HI surveys: (1) the Widefield ASKAP L-band 
Legacy All-sky Blind surveY (WALLABY) is a shallow 3$\pi$ sr survey ($z<0.26$) which will survey 
the mass and dynamics of more than $6\times 10^5$ galaxies; (2) a deeper small-area HI survey, called Deep Investigation of Neutral Gas Origins (DINGO), will trace the evolution of HI ($z<0.43$) over a 
cosmological volume of $4\times 10^7$ Mpc$^3$ detecting up to $10^5$ galaxies.

\subsection{Survey design}

All large surveys, both stellar and extragalactic, struggle to 
defend the survey size with any statistical rigour. 
This is especially true for galaxy evolution studies since many free parameters are needed 
to capture the microphysics in galaxies (Fig.~\ref{f:mo1}) and the relative importance
of different microphysics may vary with the underlying density field and the galaxy's mass.

Our initial study indicates that $10^5$ galaxies over a contiguous volume is of order the size we
need to probe (i) variations on local overdensity; (ii) different redshifts; (iii) a range of galaxy 
masses at each redshift; (iv) galaxy orientations; (v) sufficient cell division in the galaxy colour-magnitude 
diagram. We also need sufficient numbers of objects per bin to counteract complexity and fine structure 
in galaxies.  A key uncertainty is the association of galaxies with local overdensity. Statistical `crowding' 
techniques (e.g. five nearest neighbours) are intrinsically uncertain for measuring local density and indeed
there are many methods on offer (\citealt{Muldrew12}). The target density needs to be sufficiently high, as 
for the SAMI Galaxy Survey, to ensure groups and clusters are well sampled (\citealt{Bryant14b}).

To examine survey parameters more qualitatively needs high-resolution cosmological simulations
where individual galaxy properties are resolved.
The necessary simulations are only now becoming available for the first time (e.g. \citealt{Schaye14}). A proper survey estimate depends ultimately on 
what is being measured. Broad categories like fast vs. slow rotators may require smaller samples 
whereas weak measurables (e.g. Gauss-Hermite parameters $h_3$, $h_4$), or detailed properties defined by many parameters, 
push us to larger samples. 

\citet{Naab13} have shown how synthetic IFS observations 
of the stars and gas can be extracted from the latest high-resolution simulations.
There are basic properties
that we would like to have, e.g. fractional variations in slow rotators with respect to fast rotators 
as a function of environment. \citet{Fogarty14} reveal that, in the SAMI survey, 
the ratio of slow to fast rotators may vary across cluster environments. Importantly,
\citet{Naab13} argue the case for good spatial sampling to measure high order moments
($h_3$, $h_4$) of the stellar kinematics which may also vary in interesting ways over
large-scale structure due to preserved kinematic subtructure.

In summary, the Hector survey will allow us to explore how galaxy properties vary across 
a contiguous volume of the universe. 
The huge database will provide information on galaxy properties, in particular, stellar and gas kinematics, the presence of nuclear activity (starburst vs. black hole), the star formation distribution, the spread of chemical abundance in both stars and gas, and so forth. 
The Vienna conference has underscored the broad community support for the
the SAMI and Manga surveys which in turn build on the early successes of the ATLAS-3D and
CALIFA surveys. Future galaxy studies will be dominated by IFS observations of cosmologically 
significant numbers of galaxies supported by large HI surveys.

\section*{Acknowledgements}
\noindent
JBH acknowledges an ARC Federation Fellowship and LIEF grant for the SAMI 
instrument and an
ARC Australian Laureate Fellowship for the ongoing development of the Hector instrument. JBH
acknowledges insights from members of the SAMI team, including S. Croom, J. Bryant,
J. Allen, L. Fogarty, S. Richards and R. Sharp. We are indebted to J. Lawrence, the Anglo-Australian 
Observatory and the Anglo-Australian Telescope for their role in realising the SAMI project.
My thanks to G. Cecil for producing Fig.~\ref{f:sami} and to T. Naab for 
insights on where simulations can take us.

\end{document}